\newif\ifACM

\ifACM
\else
\fi

\ifACM
\documentclass[sigconf, anonymous, review]{acmart}

\usepackage{listings}
\usepackage{multirow}

\AtBeginDocument{%
  \providecommand\BibTeX{{%
    \normalfont B\kern-0.5em{\scshape i\kern-0.25em b}\kern-0.8em\TeX}}}

\setcopyright{acmcopyright}
\copyrightyear{2018}
\acmYear{2018}
\acmDOI{10.1145/1122445.1122456}

\acmConference[AFT 2020]{AFT 2020: ACM Advances in Financial Technologies}{October 21--23, 2020}{New York, NY}
\acmBooktitle{AFT 2020: ACM Advances in Financial Technologies, October 21--23, 2020, New York, NY}
\acmPrice{15.00}
\acmISBN{978-1-4503-XXXX-X/18/06}

\else
\documentclass[conference]{IEEEtran}

  \usepackage{cite}

\usepackage{amsfonts}  

\usepackage{amsmath}  

\usepackage{multirow}

\usepackage{hyperref}

\usepackage{graphicx}

\usepackage{listings}

\usepackage{caption}
\fi

\begin{document}
\title{Layer 2 Atomic Cross-Blockchain Function Calls}

\ifACM
\else
\author{
    \IEEEauthorblockN{
    	Peter Robinson\IEEEauthorrefmark{1}\IEEEauthorrefmark{2},
    	Raghavendra Ramesh\IEEEauthorrefmark{1} 
         } 
    \IEEEauthorblockA{\IEEEauthorrefmark{1}Protocol Engineering Group and Systems (PegaSys), ConsenSys}
    \IEEEauthorblockA{\IEEEauthorrefmark{2}School of Information Technology and Electrical Engineering, University of Queensland, Australia}
    \IEEEauthorblockA{
    	peter.robinson@consensys.net, 
    	raghavendra.ramesh@consensys.net
	}
}

\maketitle

\thispagestyle{plain}
\pagestyle{plain}

\fi

\begin{abstract}
The Layer 2 Atomic Cross-Blockchain Function Calls protocol allows composable programming across Ethereum blockchains. It allows for inter-contract and inter-blockchain function calls that are both synchronous and atomic: if one part fails, the whole call graph of function calls is rolled back. Existing atomic cross-blockchain function call protocols are Blockchain Layer 1 protocols, which require changes to the blockchain platform software to operate. Blockchain Layer 2 technologies such as the one described in this paper require no such changes. They operate on top of the infrastructure provided by the blockchain platform software. This paper introduces the protocol and a more scalable variant, provides an initial safety and liveness analysis, and presents the expected overhead of using this technology when compared to using multiple non-atomic single blockchain transactions. The overhead is analysed for three scenarios involving multiple blockchains: the Hotel and Train problem, Supply Chain with Provenance, and an Oracle. 
The protocol is shown to provide 93.8 or 186 cross-blockchain function calls per second for the Hotel and Train scenario when there are many travel agencies, for the standard and scalable variant of the protocol respectively, given the Ethereum client, Hyperledger Besu's performance of 375 tps, assuming a block period of one second, and assuming all transactions take the same amount of time to execute as the benchmark transactions.

\end{abstract}

\ifACM
\else

\begin{IEEEkeywords}
blockchain, ethereum, cross, transaction, atomic, performance
\end{IEEEkeywords}

\fi

\section{Introduction}
\label{sec:introduction}
The Layer Two Atomic Cross-Blockchain Function Calls (LTACFC) protocol is a blockchain technology that allows function calls across blockchains that either update state on all blockchains or discard state updates on all blockchains. The protocol enables applications to access information and utilise functionality that resides on one blockchain from other blockchains.

\begin{figure}[b]
  \includegraphics[width=\linewidth]{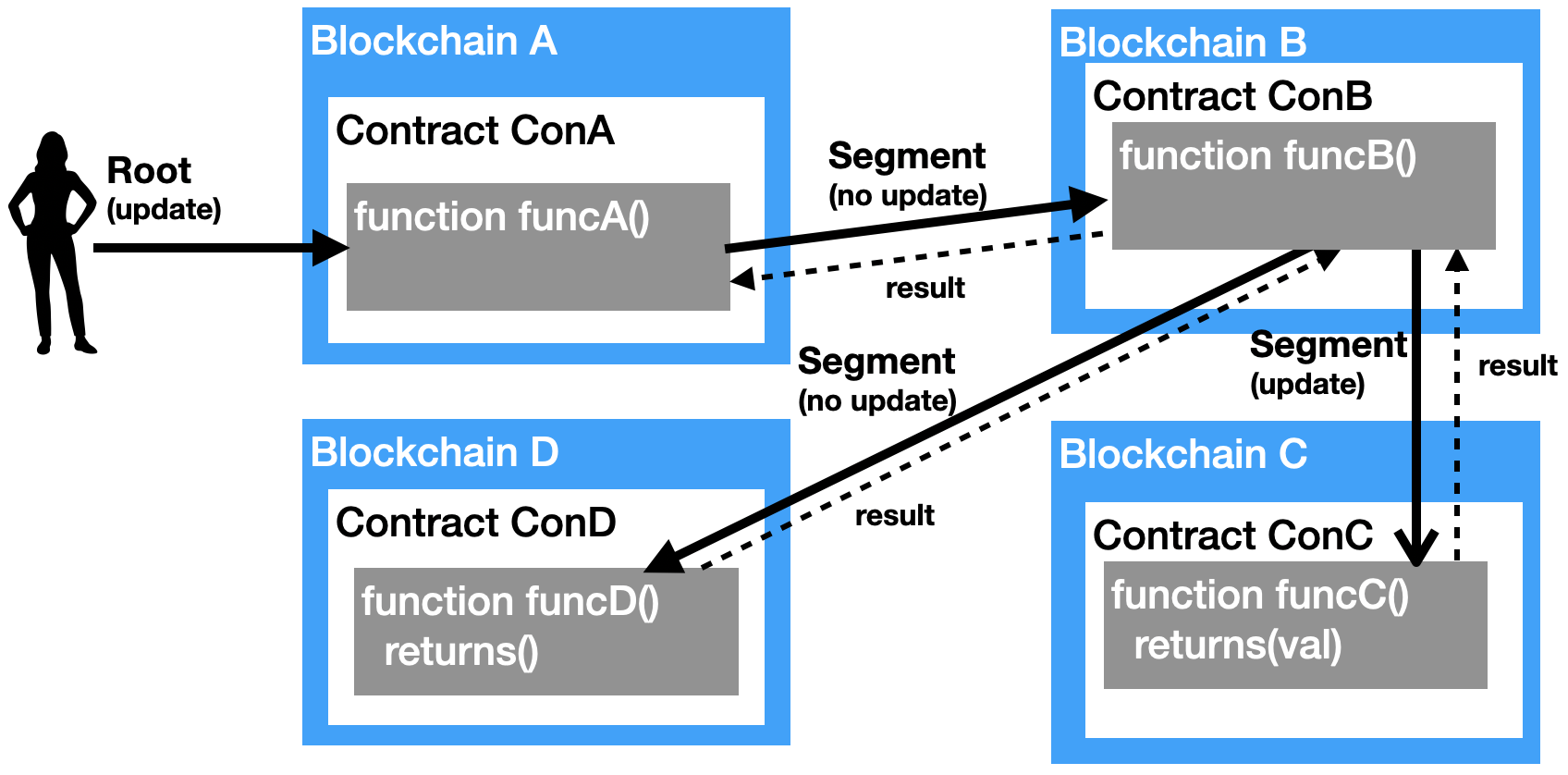}
  \caption{Cross-Blockchain Function Calls}
  \label{fig:usage}
\end{figure}

Figure~\ref{fig:usage} shows a logical representation of a cross-blockchain call graph using the LTACFC protocol. An application creates a cross-blockchain function call that goes across four blockchains. The Root Transaction executes function \texttt{funcA} in contract \texttt{ConA} on \texttt{Blockchain A}, the Root Blockchain. This function calls function \texttt{funcB} in contract \texttt{ConB} on \texttt{Blockchain B}, that in turn calls functions \texttt{funcC} and \texttt{funcD}. The function calls can update state on each blockchain and return values across blockchains. The atomic nature of the technology ensures that either all state updates are committed or all are discarded.

The LTACFC protocol can be used to create applications to solve a wide variety of problems. Three common example problems involving transactions reading and updating across multiple blockchains are presented so that the technology can be evaluated. The \textit{Hotel and Train} problem has a travel agency book a hotel room and a train seat. This requires the state of three blockchains to be updated atomically. The \textit{Supply Chain Provenance} problem allows for selective transparency of supply chain events between a Supply Chain blockchain and a Provenance blockchain. This requires two blockchains to update atomically. The \textit{Oracle} problem involves using the result of a function call on one blockchain, for example to read the current price of a commodity, in a function call on another blockchain. 

The protocol has initially been targeted at Ethereum blockchains. However, the protocol should work equally well on any blockchain that supports smart contracts. As such, this protocol can be considered a cross-blockchain protocol for heterogeneous blockchain communications. 

The protocol combines the ideas of Block Header Transfer~\cite{robinson-consensus-crosschain} for cross-blockchain consensus from BTC Relay~\cite{btc-relay}, Clearmatics' Ion project~\cite{Clearmatics2018c}, and Cosmos Inter-Blockchain Communications (IBC)~\cite{cosmos2016}, with the ideas for coordinating transactions and contract locking from Atomic Crosschain Transactions~\cite{crosschainwhitepaper, robinson2019b} and Atomic Cross Shard Function Calls using System Events and Live Parameter Checking~\cite{robinson-cross-shard} to provide an atomic synchronous composable cross-blockchain function call technology. 

\begin{figure}
  \includegraphics[width=\linewidth]{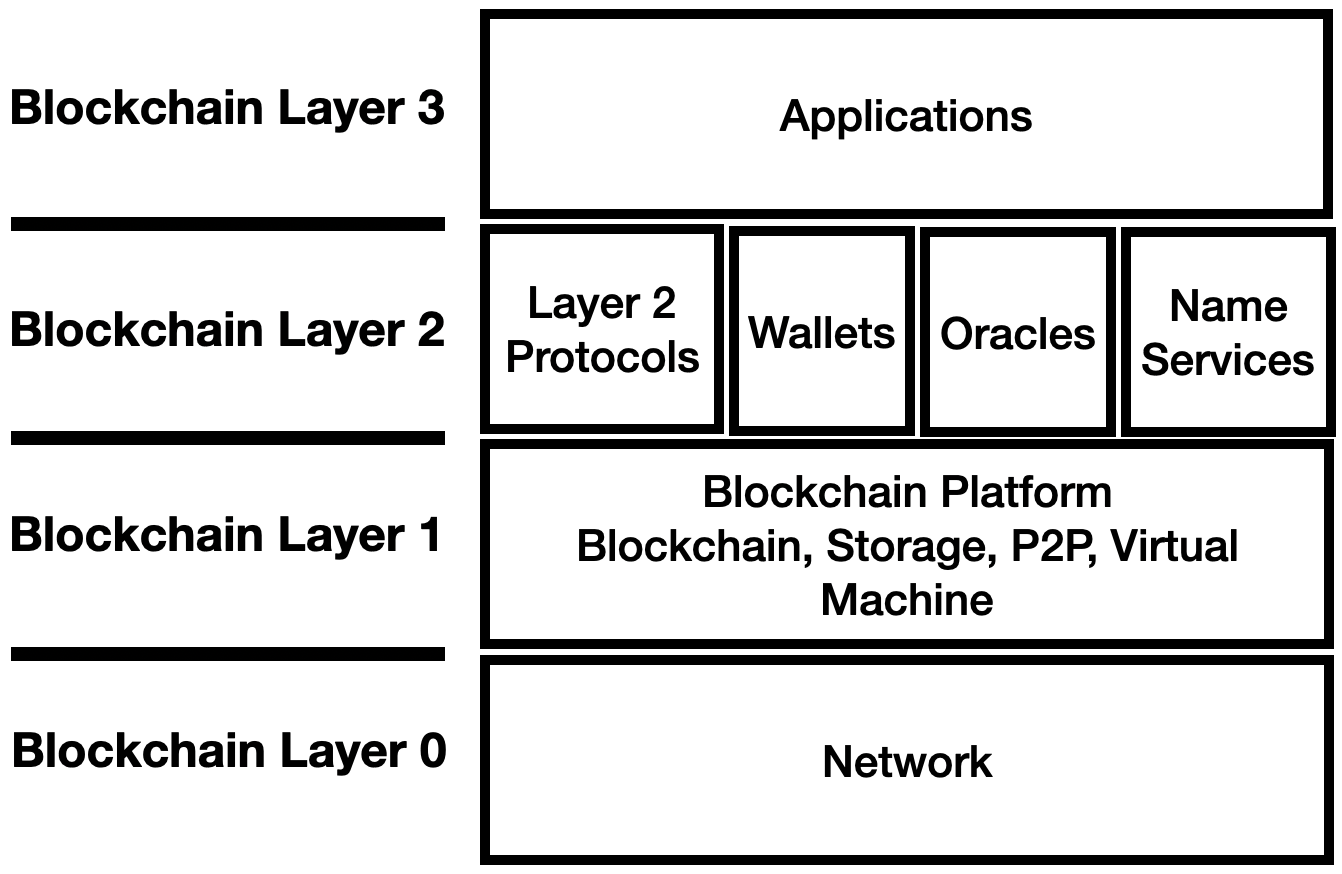}
  \caption{Blockchain Layer Architecture}
  \label{fig:layers}
\end{figure}

There are multiple similar but disparate definitions of the layers of the blockchain protocol stack~\cite{enteth20, layer2, layer2a}. Figure~\ref{fig:layers} shows the naming convention used in this paper. Blockchain Layer 0 is the underlying network over which nodes in the blockchain platform communicate. This network is typically the Internet, though could be an intranet or a Supervisory Control and Data Acquisition (SCADA) network. Blockchain Layer 1 is the blockchain platform layer. This layer includes the blockchain, state storage, peer-to-peer protocol, and the virtual machine in which contract code executes. Blockchain Layers 2 and 3 consists of contract code that executes in the virtual machine and associated off-blockchain software. Blockchain Layer 2 contracts can be Oracles, Wallets, Name Services such as Ethereum Name Service (ENS)~\cite{ens2018} or Blockchain Layer 2 protocol code.  Blockchain Layer 3 consists of application code. 

Blockchain Layer 2 protocols do not change the blockchain platform software to operate. From the perspective of blockchain platforms, they act as applications. They can utilise contract code and servers external to the blockchain to operate. Blockchain Layer 2 protocols are appropriate for situations in which users are unwilling or unable to modify their blockchain platform software.

Creating a cross-blockchain protocol that provides atomic behaviour is a complex problem~\cite{crosschain-deals}. Creating a cross-blockchain protocol that does not require changes to the blockchain platform software is significantly more challenging. This paper introduces a new Blockchain Layer 2 protocol that provides atomic cross-blockchain function calls, presents an initial security analysis, and presents the expected overhead of using this technology when compared to using non-atomic single blockchain transactions. A methodology for using this technology to provide a scalable solution involving many blockchains is presented. Additionally, this paper explains how this technology could be modified for use as an Ethereum 2 Cross Shard function call technique.

The LTACFC source code is available on github.com\cite{ltacfc_github}. The code is open source and available under an Apache 2 license.

This paper is organised as follows: the \textit{Background} section introduces Block Header Transfer for cross-blockchain consensus and introduces Ethereum. Next the \textit{LTACFC Protocol} section describes the protocol. The \textit{Performance Overhead} section describes the expected overhead of using this protocol over using separate single blockchain transactions. The \textit{Security Analysis} section analyses the Safety and Liveness of the protocol. The \textit{Scenarios} section explains the Hotel and Train, Supply Chain Provenance, and Oracle usage scenarios. The \textit{Experimental Set-up} explains how results have been gathered. The \textit{Results} section analyses the usage scenarios given the performance overhead, and experimental results. Finally, the \textit{Applicability to Ethereum 2 Cross Shard Function Calls} section explains the applicability of this protocol to Ethereum 2.

\section{Background}
\subsection{Block Header Transfer}
\label{ref:bhtransfer}
\begin{figure}
  \includegraphics[width=\linewidth]{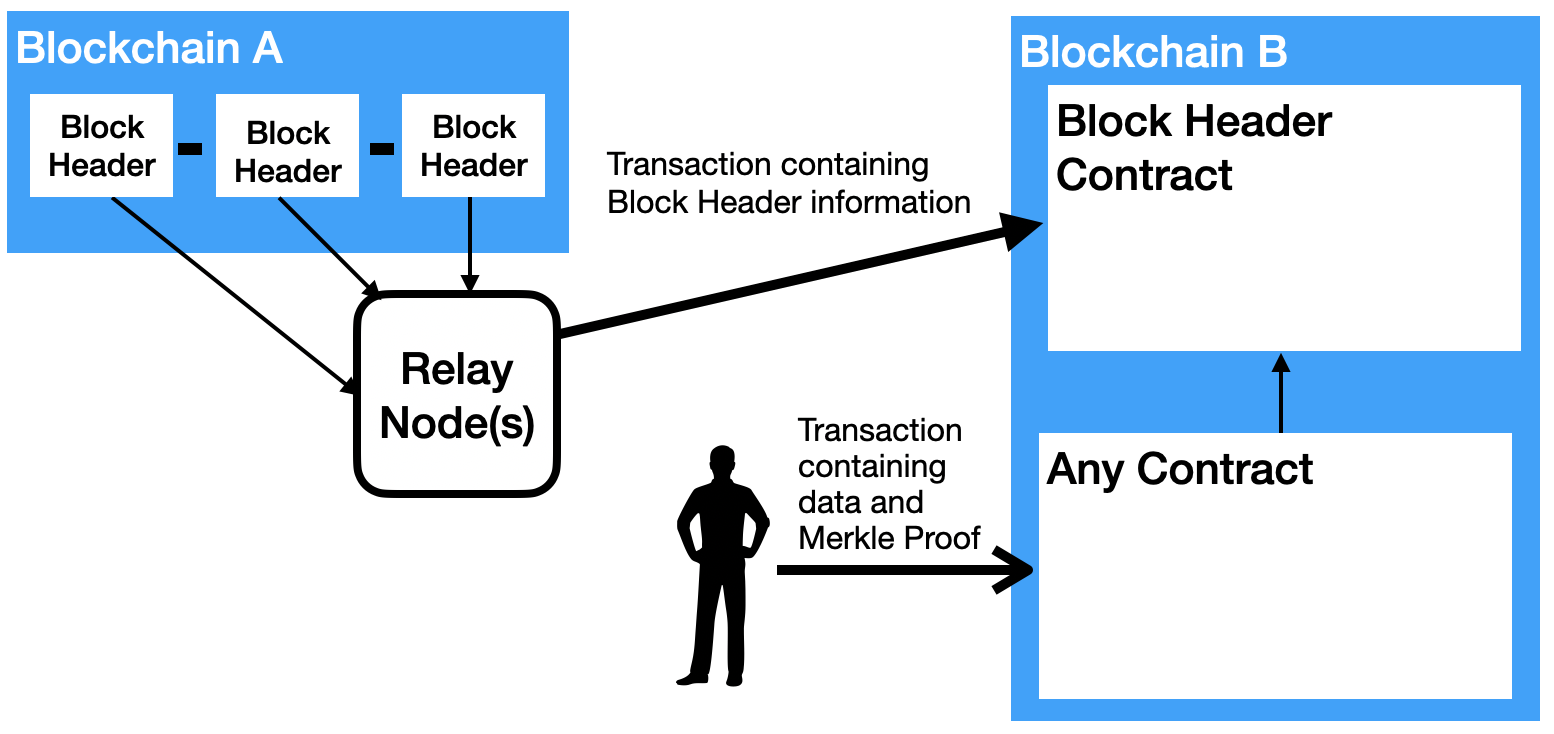}
  \caption{Block Header Transfer}
  \label{fig:blockheadertransfer}
\end{figure}

Block Header Transfer~\cite{robinson-consensus-crosschain} techniques rely on Relay Nodes reading block headers from a source blockchain and submitting the block headers to a Block Header Contract on a destination blockchain, as shown in Figure~\ref{fig:blockheadertransfer}. 

The methodology for the Block Header Contract trusting the block headers is dependent on the particular technique. For instance, with BTC Relay~\cite{btc-relay}, trust is derived from the Proof of Work (PoW) difficulty of the Bitcoin blockchain in combination with a reliance on multiple Relay Nodes being prepared to present blocks from the best canonical chain. The \textit{root of trust} is an initial trusted block header, which could be the genesis block. With Clearmatics' Ion project and Cosmos IBC the block headers must be signed by a threshold number of trusted validators. In these cases the root of trust is the set of public keys that are registered with the Block Header Contract, one for each validator.

Transactions that leverage the block headers can be submitted to the destination blockchain once the block headers are trusted. The transactions contain a block hash, Merkle Proof~\cite{merkle-tree}, and data. The data, in combination with the Merkle Proof and the block header that matches the block hash prove that the data was part of the block, and hence should be trusted. The code in \texttt{Any Contract} in Figure~\ref{fig:blockheadertransfer} can execute logic based on the fact that the data has been proven to have been included in a block on the source blockchain.

An advantage of this type of scheme is that multiple transactions that leverage the one block header can be submitted to the destination blockchain. The complex cross-blockchain consensus only has to occur once for all transactions that relate to the block. However, some schemes, such as BTC Relay, rely on all block headers being transferred, even if no transactions will use them. An efficient scheme needs to be able to determine which blockchains need which block headers, and only transfer block headers that are needed. 

\subsection{Ethereum}
\label{ref:ethereum}
Ethereum~\cite{wood2016a} is a blockchain platform that allows users to deploy and execute computer programs known as Smart Contracts. Smart Contracts are typically written in the Solidity~\cite{solidity} programming language. Source code is compiled into a bytecode representation. The bytecode is deployed to the distributed ledger using a contract creation transaction. 

Ethereum \textit{transactions} update the state of the distributed ledger, do not return values, and can emit log information. In addition to contract deployment transactions, transactions are also used to call functions in the Smart Contracts and to transfer Ether, the native coin of Ethereum, between accounts. \textit{View} function calls can be executed on the Smart Contract code. These View function calls return a value and do not update the state of the Smart Contract. 

Blocks consist of groups of transactions. Ethereum nodes come to consensus to determine the next block that will become part of the chain of blocks. As such, execution of transactions involves nodes across the network. In contrast, View function calls execute on a single node using the node's local copy of the distributed ledger. 

Ethereum MainNet, the most widely used public Ethereum network, uses a Proof of Work (PoW)~\cite{nakamoto2008, wood2016a} consensus algorithm. Ethereum 2, due to go live incrementally starting in 2020, will use a Proof of Stake (PoS) \cite{ethereum2-beacon-chain} consensus algorithm. Ethereum is also deployed in permissioned consortium networks~\cite{enteth20}. In these deployments Proof of Authority (PoA) consensus algorithms such as Istanbul Fault Byzantine Tolerant (IBFT)~\cite{ibft} and Istanbul Fault Byzantine Tolerant version 2 (IBFT2)~\cite{ibft2} are used. 

A block is deemed \textit{final} when it can no longer be changed. All transactions contained within a finalised block are also deemed final. PoW consensus provides probabilistic finality~\cite{robinson2019c}. As more blocks are added to the end of a PoW consensus blockchain, the probability of old blocks being removed due to a heavier chain being found decreases. PoS as implemented in Ethereum 2 provides a combination of probabilistic finality and fixed finality. All blocks are deemed final on the first block of a new 32 block epoch. Within epochs, blocks have probabilistic finality, depending on the number of blocks added to the chain. PoA consensus blockchains using IBFT2 offer instant finality~\cite{ibft2}: a block is final as soon as it is added to the blockchain.

\subsection{Ethereum Transaction Receipts and Events}
\label{ref:receipts}
Transactions in Ethereum can generate log events programmatically. This information is stored as logs in transaction receipts. The transaction receipts for all transactions in a block are stored in a modified Merkle Patricia Tree. The root hash of this tree is stored in the Ethereum block header. The log event information includes: the address of the contract that emitted the event, an identifier known as a \textit{topic} that specifies the type of event that is emitted, and a data blob containing the encoded event parameters.

This ability to programmatically produce events can be used to produce information on the source blockchain that can be consumed on the destination blockchain. This is the technique that Clearmatics' Ion project uses. It allows contracts on the destination blockchain to effectively read from the source blockchain.

\section{LTACFC Protocol}
\label{sec:methodology}
\subsection{Cross-Blockchain Consensus}
\label{sec:consensus}
\begin{figure}
  \includegraphics[width=\linewidth]{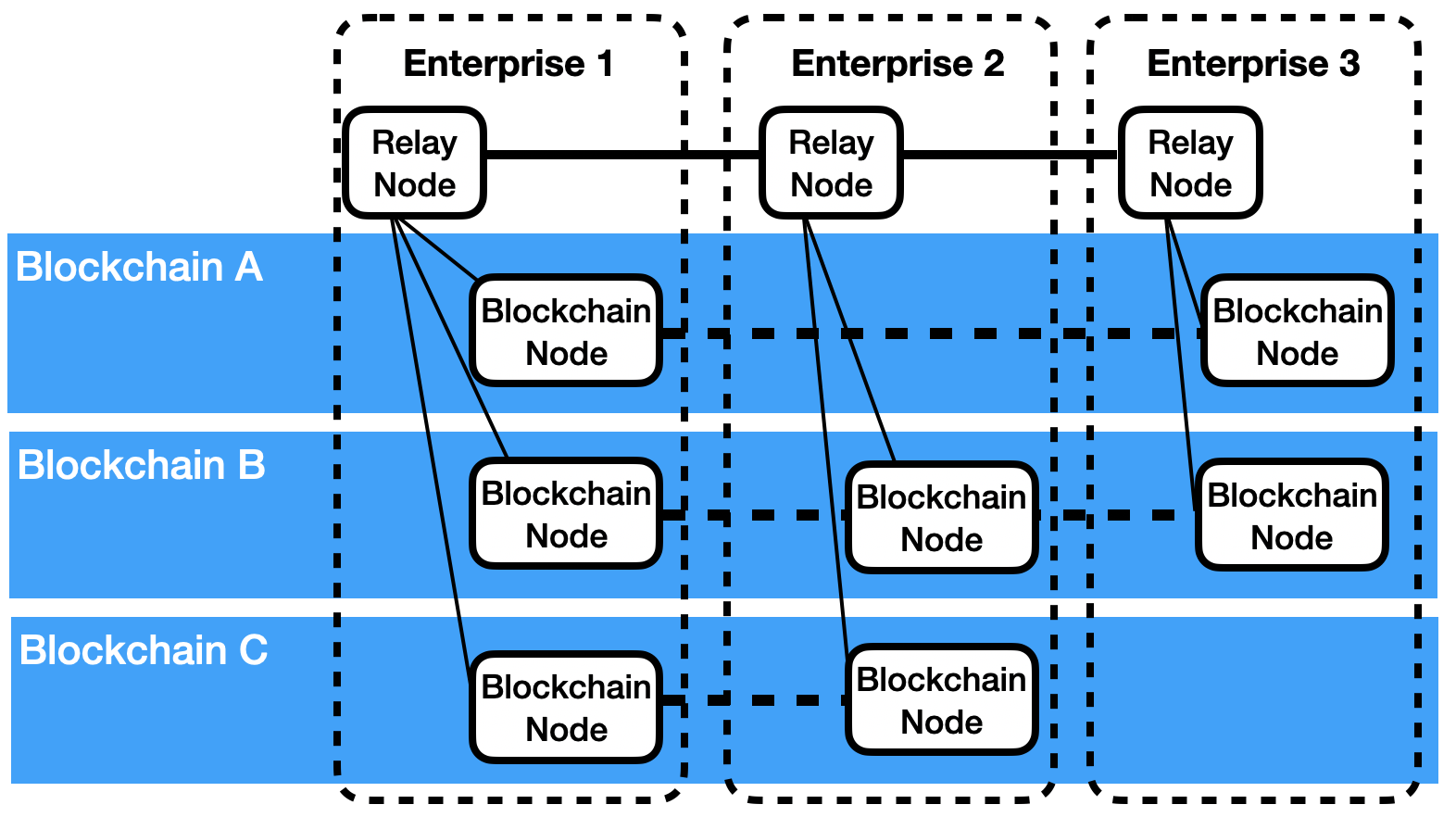}
  \caption{Example Deployment Architecture}
  \label{fig:network}
\end{figure}

Figure~\ref{fig:network} shows an example deployment architecture for companies using the LTACFC protocol. Multiple companies could host blockchain nodes on a variety of consortium or public blockchains. They could also host Relay Nodes to communicate block headers between blockchains. An important aspect of the deployment scenario is that Enterprise 2 does not have any blockchain nodes on Blockchain A, and Enterprise 3 does not have any blockchain nodes on Blockchain C. The cross-blockchain consensus mechanism must give Enterprise 2 and 3 enough information about transactions that have occurred on the blockchains they are not part of to be certain that those transactions have been included in the respective blockchains.

There are three alternative methodologies for transferring signed block headers. The appropriate methodology is deployment dependant. The first approach is that each enterprise signs a block header and submits the block header along with their signature in a transaction to the destination blockchain. There are two issues with this approach. Firstly, if there is a lack of crossover between blockchains, then only a small number of enterprises would sign block headers from one blockchain when destined for another blockchain. Consequently, the destination blockchain's trust in the validity of the block headers from the source chain will be reduced. A second issue with this approach is that a transaction needs to be put onto the destination blockchain for each relay node. This will impact the performance of the destination blockchain as one transaction per enterprise will be submitted to the blockchain, to communicate the block header and the enterprise's signature.

The second approach is that each enterprise could broadcast signed block headers to all other Relay Nodes that belong to enterprises that have nodes on the blockchain. Each Relay Node would gather signatures for a block header. Enterprises that have transactions that need the block header to be transferred from one blockchain to another could submit a transaction containing the multiply-signed block header. Subsequent transactions to add the same block header would fail. An issue with this approach is that $N \times (N - 1)$ messages need to be communicated between Relay Nodes, where $N$ is the number of Relay Nodes for a blockchain.

A final approach is to request that block headers are signed when needed. When an enterprise needs a block header transferred to a blockchain, it requests all other Relay Nodes on a blockchain to sign a block header. They gather the responses and submit the multiply signed block header in a transaction to destination blockchains. Subsequent transactions to add the same block header would fail. This approach would result in $2 \times N \times (N-1)$ messages if all enterprises simultaneously needed a block header signed. However, this would be a good approach when block header transfer was only needed intermittently.

For the second and third alternative, rather than using Elliptic Curve Digital Signature Algorithm (ECDSA) signatures, Boneh-Lynn-Shacham (BLS) threshold signatures~\cite{bls2004,bls-threshold,bls-threshold-youtube} could be used. This would have the advantages that the signature would be small, and the number of signers and the identity of signers who signed the message could be kept secret. 

\subsection{Cross-Blockchain Control Contract}
Cross-Blockchain Control Contracts are used to manage parts of a cross-blockchain function call. An instance of this contract is deployed to each blockchain that will participate in cross-blockchain function calls. The address of the contract and the blockchain identifier of the blockchain are registered with Block Header Contracts on all blockchains. This allows events emitted by this contract to be trusted as being from a Cross-Blockchain Control Contract. The following sections describe the main functions of the contract. 
\begin{figure*}
  \includegraphics[width=\linewidth]{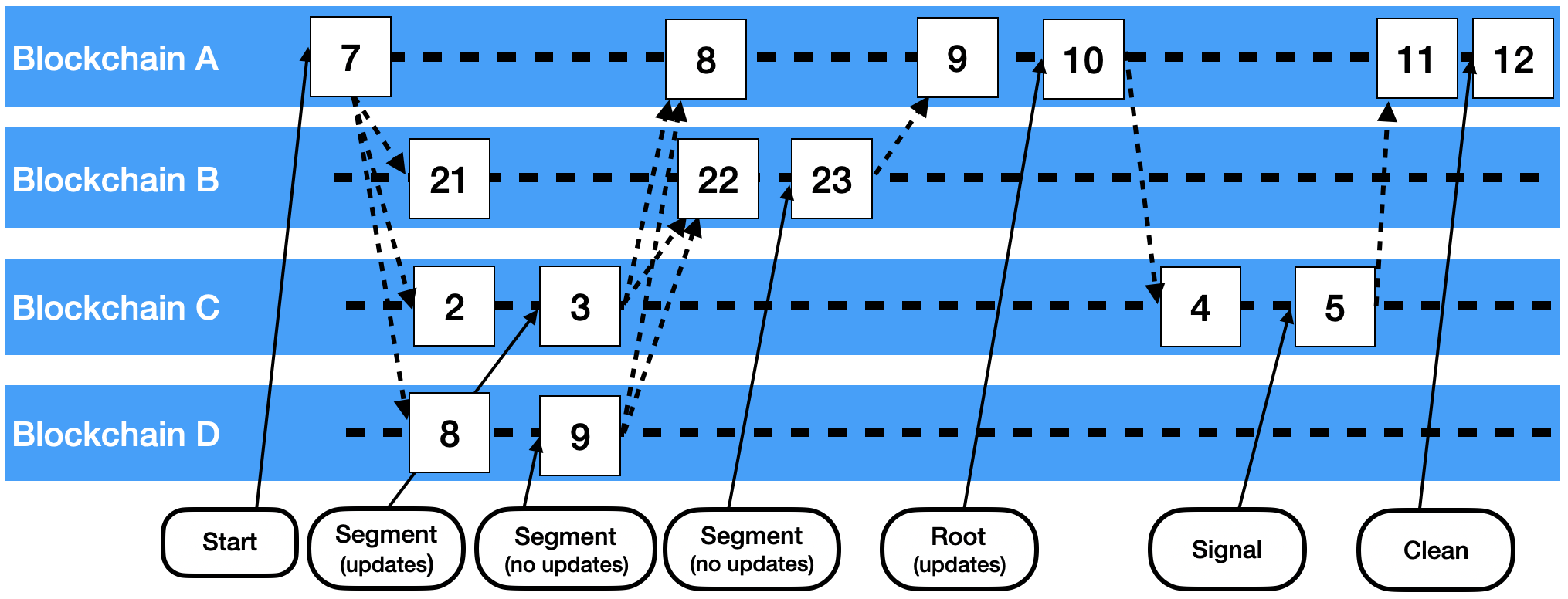}
  \caption{Sequence of Function Calls}
  \label{fig:sequence}
\end{figure*}

\subsubsection{Start}
The \textit{Start} function is called on the blockchain which contains the entry point function call for the cross-blockchain function call, the Root Blockchain. The function registers the account that will submit all of the transactions on all of the blockchains, the call graph of the cross-blockchain function call including expected function parameter values, a time-out in terms of block time stamp of the Root Blockchain, and a cross-blockchain transaction identifier. This information is emitted in an event know as the \textit{Start Event}. The transaction that submits the Start function call is known as the \textit{Start Transaction}.

\subsubsection{Segment}
The \textit{Segment} function is called to request a function on a contract be called as part of a cross-blockchain function call. The Start Event, the block hash of the block containing the Start Transaction, a Merkle Proof proving the Start Event is related to the block, and an indicator of where the function call lies in the call graph are submitted as parameters to prove that this function is part of the cross-blockchain function call. The code is able to verify that the latest block it has received from the Root Blockchain has a block time stamp older than the time-out value. Additionally, a set of \textit{Segment Events} containing function call return results, with block hashes, and Merkle Proofs are passed in to prove that subordinate function calls on other blockchains have been called and have returned certain results. These function call results match function calls that are described in the Start Event. 

When the application contract code executes, if a cross-blockchain function call is encountered, the application code calls a function \textit{CrossCall} in the Cross-Blockchain Control Contract, passing in the actual parameters of the function call. The  CrossCall function compares the actual parameter values with the expected parameter values as specified in the Start Event. The function returns results that have been indicated should be returned in Segment Event for the function call. Section \ref{sec:locking} describes contract locking, which is used if state updates occur.

A Segment Event is emitted to publish the return result or error result of the function, and the list of locked contracts. The transactions that submit the Segment function calls are known as \textit{Segment Transactions} and the blockchains they execute on are termed \textit{Segment Blockchains}.

\subsubsection{Root}
The \textit{Root} function is called on the Root Blockchain to call the entry point function call for a cross-blockchain function call. The Root function call has similar parameters to the Segment function call, taking a Start Event and a set of Segment Events, along with the block hashes and Merkle Proofs to allow the event contents to be validated. Similar to the Segment function call, expected and actual parameter checking is completed along with contract locking.

If a Root function call completes successfully, any locked contracts on the Root Blockchain are unlocked and provisional state updates are committed. A \textit{Root Event} is emitted indicating that all provisional updates on blockchains should be committed. The Root function emits a Root Event indicating that all updates on all other blockchains should be discarded if any of the Segment functions returned error results or if an error occurs while executing the entry point function call. 

If the time stamp of the most recent block on the Root Blockchain is after time-out, then any account can submit a transaction that calls the Root function to cancel the cross-blockchain function call. In this situation a Root Event indicating that all updates on all other blockchains should be discarded is emitted.

\subsubsection{Signal}
The \textit{Signal} function is called on blockchains that have updates that need to be committed or ignored. The Start Event, Root Event, and Segment Event for the blockchain, along with block hashes and Merkle Proofs are passed in as parameters. Calling the function requests that contracts on a blockchain locked due to a state update due to a Segment function call be unlocked and the state updates be either committed or ignored depending on the information in the Root Event. A Signal Event is emitted to indicate that the contract has been unlocked.

\subsubsection{Clean} 
The \textit{Clean} function is called on the Root Blockchain to remove the cross-blockchain identifier from the Cross-Blockchain Control Contract, thus terminating the cross-blockchain function call. The Start Event, all Segment Events, and all Signal Events need to be submitted along with proofs, so that the control contract can be sure that all contracts that were locked due to the cross-blockchain function call have been unlocked prior to removing the cross-blockchain identifier.

\subsection{Call Graphs and Reverse Order Execution}
Figure~\ref{fig:usage} shows a logical representation of a cross-blockchain call in which an application creates a cross-blockchain function call that goes across four blockchains. The set of Cross-Blockchain Control Contract function calls that should be executed to affect this is shown in Figure~\ref{fig:sequence}.

Walking through the call sequence diagram in Figure~\ref{fig:sequence}:
\begin{enumerate}
\item A Start Transaction is submitted to blockchain A, causing the Start Event to be emitted. The block header that the transaction is included in is transferred to all blockchains that are part of the call graph.
\item Segment Transactions are submitted to blockchains C and D. These execute the leaf parts of the call graph. Segment Events are emitted. The block headers for blocks that these transactions are included in are transferred to the blockchains that they will be needed on.
\item A Segment Transaction is submitted to blockchain B. This executes the call next up the call graph from the leaves. A Segment Event is emitted. The block header for the block that this transaction is included in is transferred to blockchain A.
\item The Root Transaction is submitted to blockchain A. This executes the root part of the call graph and emits a Root Event. The block header for the block containing the transaction is transferred to the blockchain that had a state update: blockchain C.
\item A Signal Transaction is submitted to blockchain C to commit the provisional state update. A Signal Event is emitted. The block header for the block that this transaction is included in is transferred to the blockchain that it will be needed on: blockchain A.
\item A Clean Transaction is submitted to blockchain A to terminate the cross-blockchain function call.
\end{enumerate}

It should be noted that Figure~\ref{fig:sequence} makes some assumptions:
\begin{itemize}
\item The diagram only shows blocks being produced when transactions are present. Many consensus algorithms produce blocks irrespective of whether there are transactions present.
\item The diagram shows block production times synchronised across blockchains. This is typically not the case.
\item It has been assumed an instant finality consensus algorithm such as IBFT2~\cite{ibft2} has been used on all blockchains. Block headers can only be transferred once a block is deemed final. 
\end{itemize}

\subsection{Contract Call Flow}
\begin{figure}
  \includegraphics[width=\linewidth]{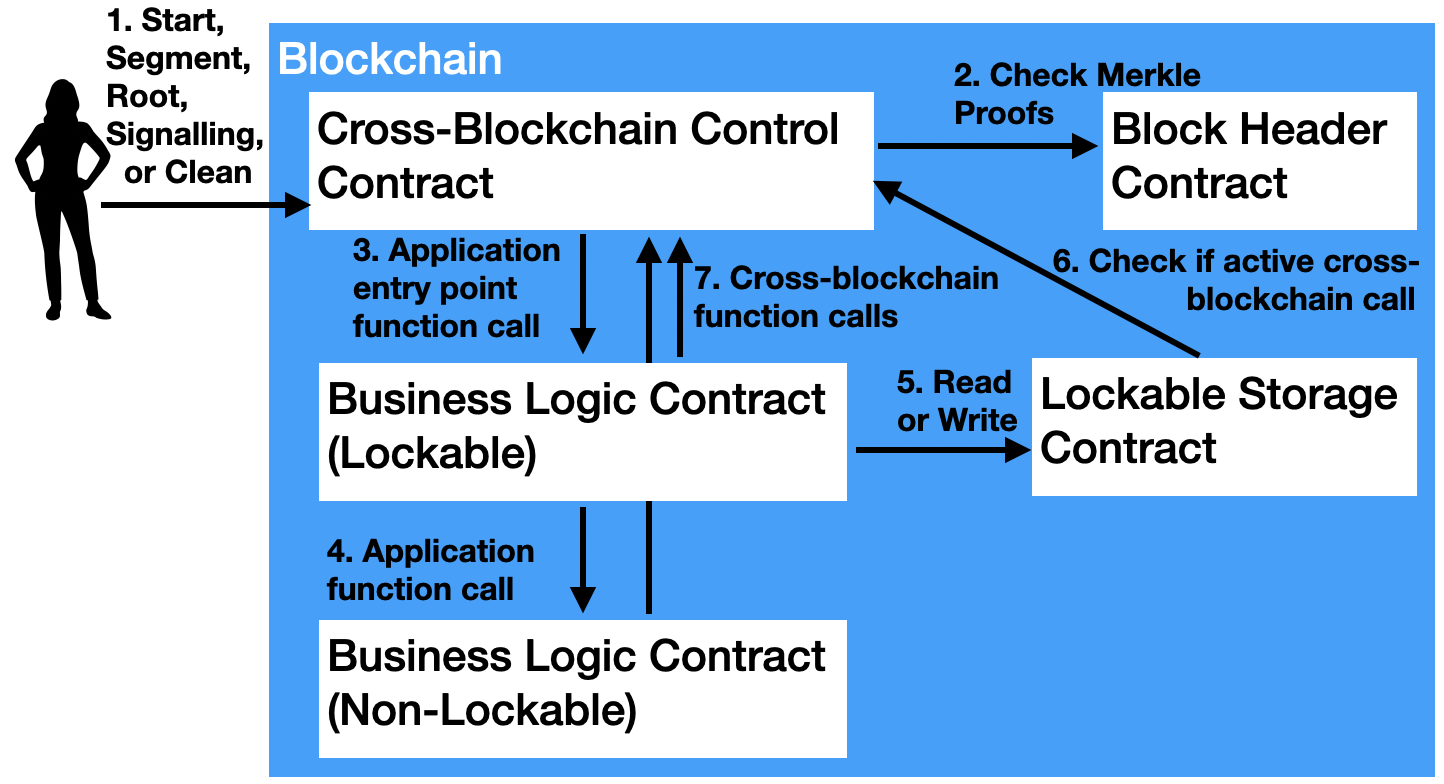}
  \caption{Contract Call Flow}
  \label{fig:calls}
\end{figure}

Figure~\ref{fig:calls} shows the call flow between contracts to facilitate LTACFCs. Walking through the steps in the diagram:
\begin{enumerate}
\item The application submits Start, Segment, Root, Signal, and Clean function calls to the Cross-Blockchain Control Contract. 
\item The Segment, Root, Signal, and Clean function call parameters include Merkle Proofs to prove events from other blockchains are valid. The Block Header Contract is used to check the validity of this event information. 
\item The Segment and Root function calls execute functions on business logic contracts. 
\item These business logic contracts may in turn call other business logic contracts. 
\item If a business logic contract can be updated as part of a cross-blockchain transaction, it will have a Lockable Storage Contract associated with it. This contract will hold all state that is modifiable during a cross-blockchain call. 
\item As part of the locking mechanism, Lockable Storage Contracts need to check with the Cross-Blockchain Control Contract to check whether a read or write is happening in the context of a cross-blockchain function call or not. 
\item Business logic contracts may also include cross-blockchain calls. To check that there is linkage between Root and Segment calls, the business logic contracts pass the function calls to the Cross-Blockchain Control Contract to check that the actual blockchain id, contract address and parameters for the cross-blockchain function call match the expected values from the Start Event.
\end{enumerate}

\subsection{Locking}
\label{sec:locking}
Lockable business logic contracts hold the parts of their data that can be updated in a cross-blockchain function call in Lockable Storage Contracts. These contracts hold state updates due to cross-blockchain calls as provisional state updates. They commit or discard the provisional updates based on Signal Transactions. The algorithms for processing read requests is shown in Listing~\ref{listing:read}, write requests is shown in Listing~\ref{listing:write}, and signalling requests is shown in Listing~\ref{listing:signal}.

\begin{lstlisting}[
%  frame=single,
  basicstyle=\footnotesize\ttfamily,
  numbers=left,
stepnumber=1, 
  firstnumber=1,
  numberfirstline=true,
  numbersep=5pt,    
  xleftmargin=0.5cm,
  morekeywords={msg},
  label=listing:read,
  caption=Locking: Read Request Processing
]
If caller is not Business Logic Contract {
  throw an error
}
Check Cross-Blockchain Control Contract:
 is there an active cross-blockchain call?
If not (normal single blockchain call) {
   If locked {throw an error} 
   Else (not locked) {Read from normal storage}
} 
Else (this is a cross-blockchain call) {
  If locked {
    Check Cross-Blockchain Control Contract:
    has this cross-blockchain call previously 
    locked the contract?
    If no {throw an error}
    If yes {Allow the read. If the value isn't
      available in provisional storage, return 
      the value in normal storage.}
   }
   Else (not locked) {Read from normal storage}
}
\end{lstlisting}

\begin{lstlisting}[
%  frame=single,
  basicstyle=\footnotesize\ttfamily,
  numbers=left,
stepnumber=1, 
  firstnumber=1,
  numberfirstline=true,
  numbersep=5pt,    
  xleftmargin=0.5cm,
  morekeywords={msg},
  label=listing:write,
  caption=Locking: Write Request Processing
]
If caller is not Business Logic Contract {
  throw an error
}
Check Cross-Blockchain Control Contract:
 is there an active cross-blockchain call?
If not (normal single blockchain call) {
   If locked {throw an error} 
   Else (not locked) {Write to normal storage}
} 
Else (this is a cross-blockchain call) {
  If locked {
    Check Cross-Blockchain Control Contract:
    has this cross-blockchain call previously 
    locked the contract?
    If no {throw an error}
    If yes {Write to provisional storage}
  }
  Else (not locked) {
    Lock the contract.
    Indicate in the Cross-Blockchain Control 
      Contract that this call is locking the 
      Lockable Storage contract
    Write to provisional storage
   }
}
\end{lstlisting}

\begin{lstlisting}[
%  frame=single,
  basicstyle=\footnotesize\ttfamily,
  numbers=left,
stepnumber=1, 
  firstnumber=1,
  numberfirstline=true,
  numbersep=5pt,    
  xleftmargin=0.5cm,
  morekeywords={msg},
  label=listing:signal,
  caption=Locking: Signal Request Processing
]
If caller is not Cross-Blockchain Control 
Contract {
  throw an error
}
If committing updates {
  Apply updates from provisional storage
}
Else (discarding updates}
  Delete provisional storage
}
Unlock the contract
\end{lstlisting}

\subsection{Simulation to Create Call Graph}
A parameter to the Start function call is the call graph to be called including function parameter values. A code execution simulation needs to be run in the application to determine expected parameter values. The simulation will need access to values from the called blockchains to determine expected parameter values if any of the values depend on contract state. 

\subsection{Call Graph Limitation}
An enterprise can only submit cross-blockchain transactions across blockchains that they have access to. For example, in Figure~\ref{fig:network}, Enterprise 1 can submit transactions that go across Blockchains A, B, and C, but Enterprise 2 can only submit transactions that go across Blockchains B and C.

\subsection{Improved Scalability Variant}
This section presents a variation on the LTACFC protocol described thus far that allows for greater scaling, inspired by Ethereum 2 Beacon Chain's \cite{ethereum2-beacon-chain} cross linking mechanism for publishing information roots to the beacon chain and Hyperservice's Network Status Blockchain's~\cite{hyperservice} holding of cross-blockchain state in a Merkle Tree. The LTACFC requires multiply signed block headers be submitted to each blockchain that they will be needed on. If there are a large number of blockchains, then a large number of transactions on each blockchain would be consumed with submitting block headers.

\begin{figure}
  \includegraphics[width=\linewidth]{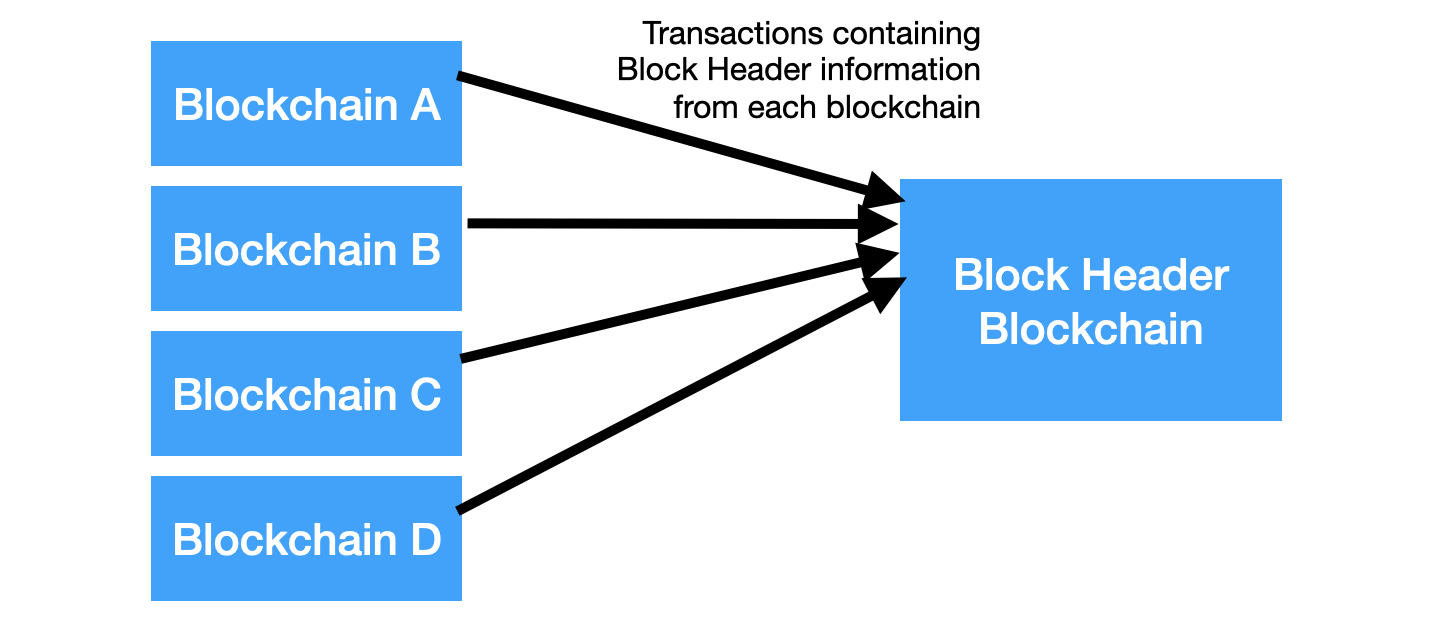}
  \caption{Submitting Block Headers to Block Header Blockchain}
  \label{fig:multi1}
\end{figure}

\begin{figure}
  \includegraphics[width=\linewidth]{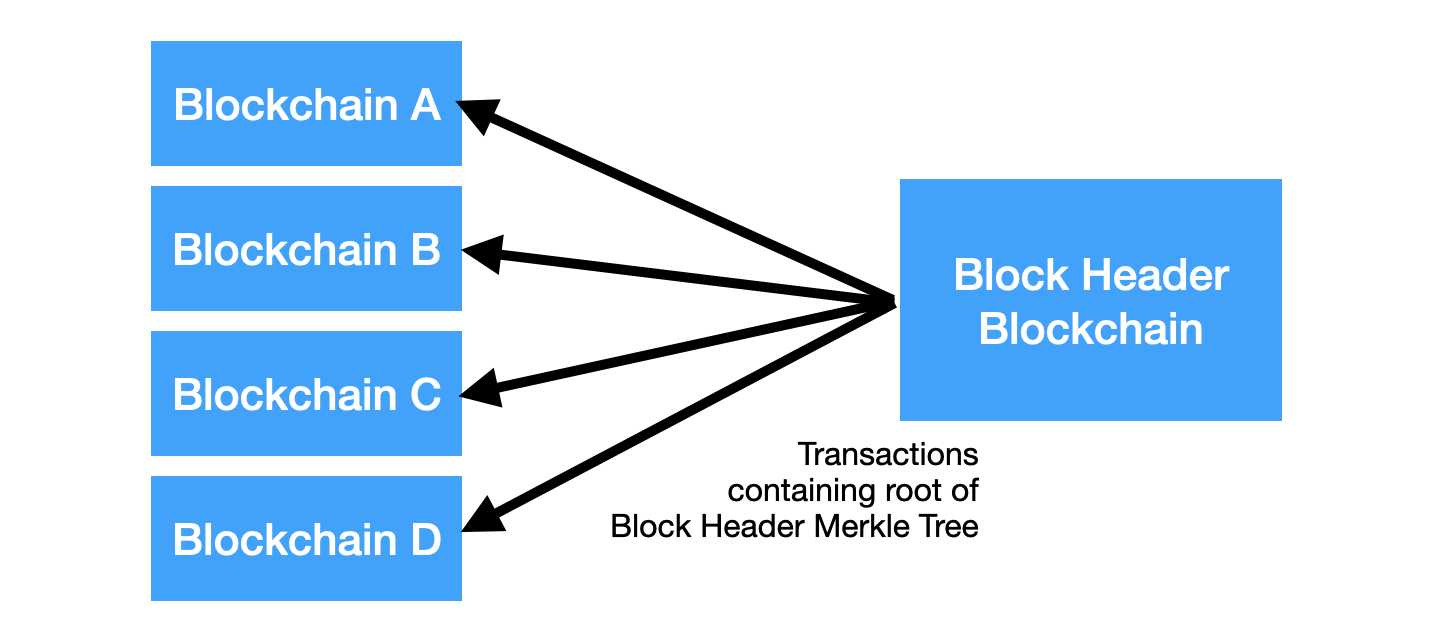}
  \caption{Submitting Root of Block Header Merkle Tree to Blockchains}
  \label{fig:multi2}
\end{figure}

A variant of the protocol is to submit block headers to a single blockchain, assemble the block headers into a Merkle Tree, and then submit the root hash of the block header Merkle Tree to all blockchains. In Figure~\ref{fig:multi1} block headers from each blockchain are submitted to a contract on the Block Header Blockchain. The message digest of block headers are constructed into a Merkle Tree as they are submitted. As shown in Figure~\ref{fig:multi2} the root of the Merkle Tree of block headers is submitted to each blockchain for each block. In this way, each blockchain only has one transaction per block to communicate a root of trust that can be used to prove values across all blockchains. 

This protocol variant provides improved scalability due to fewer transactions needing to be submitted to each blockchain to support cross-blockchain function calls. This improved scalability comes with the cost of greater latency. With the standard LTACFC protocol, a block header is transferred to a destination blockchain in one block period and then a part of a cross-blockchain function call can be submitted for the next block. With the scalability variant of the protocol, the block header is submitted to a Block Header Blockchain, and then the Merkle Root from the Blockchain Header Blockchain is submitted destination blockchain. This means that two block periods would be needed to transfer block headers with the scalability variant compared to one block period for the standard protocol.

\section{Performance Overhead}
\label{sec:perf}
This sections describes the expected impacts on performance of using this technique when compared to submitting multiple single blockchain transactions. 

\subsection{Transactions Per Second}

\begin{table}
  \centering
    \begin{tabular}{| l | l | l |}
    \hline
   \multirow{2}{*}{Blockchain} &  \multicolumn{2}{|c|}{Number of Transactions}  \\
       \cline{2-3}
                         & LTACFC Protocol & LTACFC Protocol\\ 
                         &                             & Scalability \\ 
                         &               & Variant \\ 
       \hline
Root                    & 1 (Start) & 1 (Start)  \\
Blockchain          & +1 (Root) & +1 (Root)  \\
                            & +1 (Clean) & +1 (Clean)  \\
                            & +1 for a block header & +1 for a root hash \\
                            & for each Segment & for each block \\
                            & (no updates) &   \\
                            & +2 for block headers & \\
                            & for each Segment & \\
                            & (updates)  & \\
       \hline
Segment             & 1 (Segment) & 1 (Segment)  \\
Blockchain          & +1 for a block header & +1 for a root hash \\
(no updates)       &  from the Root   & for each block \\
                            & Blockchain for the & \\
                            & Start Event & \\ 
       \hline
Segment             & 1 (Segment) & 1 (Segment)  \\
Blockchain          & +1 (Signal) & +1 (Signal)\\
(updates)            & +2 for block headers & +1 for a root hash \\
                            & from the Root & for each block \\
                            & Blockchain for the & \\
                            & Start and Root Events & \\
       \hline
  \end{tabular}
  \caption{Number of Transactions}
  \label{table:perf}
\end{table}

Table~\ref{table:perf} shows how many transactions are needed to process a cross-blockchain function call using the LTACFC technology. The number of transactions is dependant on the number of Segments, which relates to the size of the call graph, and whether the standard or the scalability variant of the LTACFC protocol is used. This is because the Root Blockchain needs to have access to the block headers for the blocks that included transactions that executed a Segment function call from each of the blockchains in the call graph, so the Clean function call knows which Segments resulted in locked contracts. For example, if the call graph shown in Figure~\ref{fig:usage} was processed then the Root Blockchain would need to process seven transactions using the standard protocol and three for the scalability variant of the protocol. Note that with the scalability variant of the protocol, the four block headers will still need to be transferred, however, the transactions will only be needed once per block for all blockchains.

The transactions used in this technology are each likely to be complex. The transactions to submit block headers will require verifying signatures. The Root, Segment, Signal, and Clean function calls involve verifying Merkle Proofs. This additional processing is likely to reduce the number of transactions that can be processed in a block.

In traditional Ethereum View calls, the return value is determined by executing a function based on an Ethereum node's local copy of the distributed ledger. With LTACFC, cross-blockchain function calls that return values to other blockchains need to included as transactions. This means that all nodes on a blockchain need to be involved in the process of executing the function call, rather than just one node.

\subsection{Transaction Size / Networking}
The Start, Root, Segment, Signal, and Clean function calls all have large parameters such as calls graphs and Merkle Proofs passed to them. It can be expected that the size of the transactions will be larger than typical transactions. Depending on the signing methodology (see Section~\ref{sec:consensus}), multiple transactions may need to be communicated between Relay Nodes to sign each block header.

\subsection{Memory}
No significant impact is expected for memory usage.

\subsection{Disk / State}
Log events are used in this technique to publish a variety of information including call graphs, function return results, and lists of locked contracts. These large log events are stored on disk.

The block headers are stored in blockchains to allow values emitted in events to be validated. The impact of this storage could be reduced by just storing the Merkle Root of the transaction receipts. Additionally, old block headers could be removed.

\section{Security Analysis}
The property of \textit{atomicity} in this protocol can be split into the following safety and liveness properties. 

It has been assumed that:
\begin{enumerate}
\item Block header transfer mechanism is trusted.
\item Root transaction will eventually (after a finite number of steps) be submitted either by the application or by others.
\item Signal transactions will eventually (after a finite number of steps) be submitted either by the application or by others.
\item Number of Byzantine nodes on each blockchain is less than the limit imposed by its enshrined consensus protocol.
\end{enumerate}

\subsection{Safety}
\textbf{Claim}: Suppose that the LTACFC cross-blockchain protocol has finished executing a cross-blockchain transaction \texttt{c}. Then:
\begin{itemize}
\item If \texttt{c} succeeds then the protocol successfully commits the state updates of all the associated (Root and Segment) transactions,
\item If \texttt{c} fails then the protocol rolls back the state updates of all the associated transactions.
\end{itemize}

\textbf{Case Analysis}: Analysed the different ways (cases) the LTACFC protocol can finish.

\textbf{Happy case}: When all transactions go through as planned before the time-out, we have that the Root Transaction commits the updated state on the Root Blockchain, and records the commit message in the Root Event. The trusted block header transfer mechanism along with the submission of Signal Transactions on subordinate blockchains, the provisional states are committed to the respective blockchains. 

\textbf{Failure of a Segment Transaction or Root Transaction}: If any Segment Transaction fails, then the corresponding Segment Event records an error. Subsequently, the Root Transaction processing reads that error, records an Ignore message in the Root Event, and discards the provisional updates of the Root Transaction. When Signal Transactions are submitted with this Root Event showing the Ignore message, then the provisional updates on subordinate blockchains are also discarded.


\textbf{Failure of the Start Transaction}: When the processing of a Start Transaction fails, the Start Event is not emitted. Without the Start Event and its Merkle Proof, the Root Transaction nor the Segment Transactions can be processed. The entire cross-blockchain transaction never gets processed. This is vacuously equivalent to a failed cross-blockchain transaction resulting in discarding of provisional states.

\textbf{Timeout}: In this scenario, we have the Start Transaction successfully processed. The Root Transaction and zero or more Segment Transactions were not processed prior to the Cross-Blockchain Timeout. In this case, from the assumption, we have that the Root Transaction is submitted eventually, and its processing records the Ignore message in the Root Event, due to the timeout. Again from the assumption, we have that the Signal Transactions are eventually processed and the provisional updates on other blockchains are discarded.

\subsection{Liveness}
\textbf{Claim}: The LTACFC protocol terminates after a finite number of steps.

\textbf{Case Analysis}: The Cross-Blockchain Timeout for a cross-blockchain transaction is recorded in the Start Event. The Start Event is included in the Root Transaction as well as other Segment Transactions. Hence, the protocol is designed to complete the transaction under consideration within the timeout period. From the assumptions we have that the Root Transaction and the Signal Transactions are submitted within a finite number of steps. Hence, the LTACFC protocol always terminates.

Note that there are no centralised entities in this protocol, apart from the relay nodes. From the assumption that the block header transfer mechanism is trusted, and from the assumption that the number of Byzantine nodes in a blockchain is less than the limit imposed by its enshrined consensus protocols, we have that the nodes becoming Byzantine does not affect the termination of the LTACFC protocol.

\subsection{Deadlocks and Livelocks}
The liveness property implies that the LTACFC protocol is deadlock free. However, the LTACFC protocol is not livelock free. Consider the case of two applications \texttt{A1} and \texttt{A2} submitting two cross-blockchain transactions \texttt{T1} and \texttt{T2}, both wanting to update the same contracts \texttt{C1} and \texttt{C2} on blockchains \texttt{B1} and \texttt{B2} respectively. Application \texttt{A1} has as the target of the Root Transaction contract \texttt{C1} and application \texttt{A2} has as the target of the Root Transaction contract \texttt{C2}. The Segment Transactions will go through, locking contract \texttt{C2} for application \texttt{A1} and contract \texttt{C1} for application \texttt{A2}. However, the Root Transactions to update contracts \texttt{C1} and \texttt{C2} will fail as the contracts would already be locked. Both cross-blockchain transactions will fail. The applications could repeatedly resubmit these transactions, and repeating the same failure situation. Hence, it is possible for the LTACFC protocol to be in a livelock continuously.

\subsection{Replay Protection}
The Start function call is called with a Cross-Blockchain Function Call Identifier. The identifier is emitted in the Start Event. This identifier and the Root Blockchain's identifier are recorded in the Root Blockchain and Segment Blockchains Cross-Blockchain Control Contracts to ensure Segment Transactions and Root Transactions aren't submitted more than once for the one cross-blockchain call.

\section{Scenarios}
\label{sec:scenarios}
\subsection{Hotel and Train}
A common example distributed transaction problem is know as the Hotel and Train problem. In this scenario, a travel agent needs to ensure the atomicity of a combined booking transaction. In other words, the travel agent needs to ensure that they either book both the hotel room and the train seat, or neither, so that they avoid the situation where a hotel room is successfully booked but the train reservation fails, or vice versa. There are three permissioned blockchains involved: the travel agency runs a blockchain, and each hotel and train travel company also maintains their own blockchain. 

\begin{figure} [h]     
\includegraphics[width=\columnwidth]{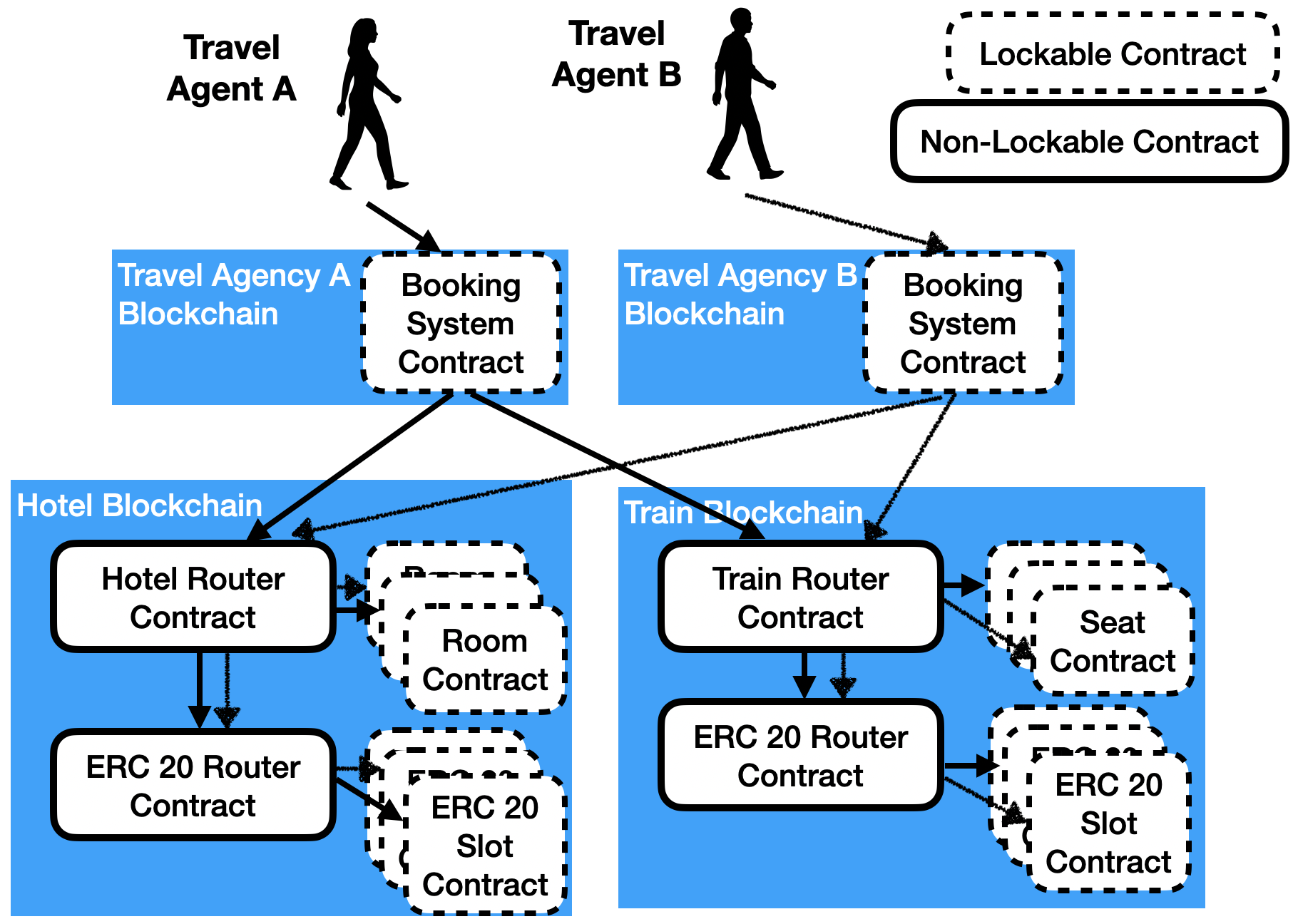}       
\caption{Hotel and Train Scenario}
\label{fig:hoteltrain}
\end{figure}

Figure~\ref{fig:hoteltrain} illustrates the implementation of a `Hotel and Train' reservation system using the LTACFC protocol. The train blockchain and the hotel blockchain host especially designed contracts that conform to a Router-Item pattern. With this pattern, a non-lockable router contract is used to access lockable item contracts. For example, when booking a train seat, a Train Router contract function is called that locates a Train Seat contract that is not locked that indicates it represents a train seat that is available for the date that the train seat needs to be reserved for. In this way multiple parallel cross-blockchain calls that need to lock contracts that contain updates can occur in parallel, thus allowing for multiple seat reservations to occur in parallel. Similarly, the ERC 20 token contracts \cite{eip20, erc20-standard} have been modified to conform to this pattern, which allows for multiple parallel payments. 

Travel agents purchase ERC 20 tokens on the hotel and train blockchains, which they can then use to pay for accommodation and travel. They book hotel rooms and train seats by executing cross-blockchain transactions, paying for the rooms and seats in the same transaction.

\subsection{Supply Chain with Provenance}
A vendor may wish to publish information to customers to provide assurances of product provenance. With only a single global blockchain all transaction details must be published and will be visible to all participants of the global blockchain. The vendor may prefer, for example, to keep the identities of its suppliers secret from competing vendors. 

\begin{figure} [h]     
\includegraphics[width=\columnwidth]{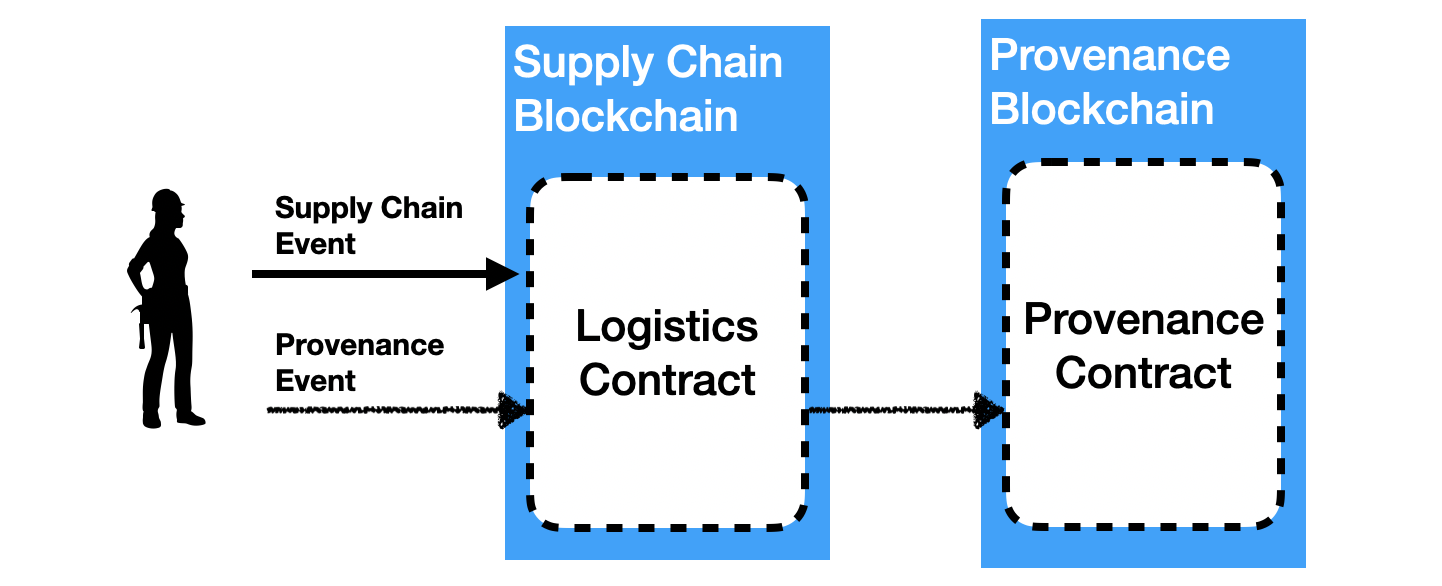}
\caption{Supply Chain with Provenance Scenario}
\label{fig:supplychain}
\end{figure}

This scenario is illustrated in Figure~\ref{fig:supplychain} in which two blockchains are used. A \textit{Supply Chain} blockchain maintains all of the transactions between a vendor and its suppliers. The \textit{Provenance} blockchain holds all the information required to assure customers of various aspects of the goods being purchased. When a provenance event occurs, a cross-blockchain transaction is used to update the Supply Chain blockchain and the Provenance blockchain. 

\subsection{Oracle}
\begin{figure} [h]     
\includegraphics[width=\columnwidth]{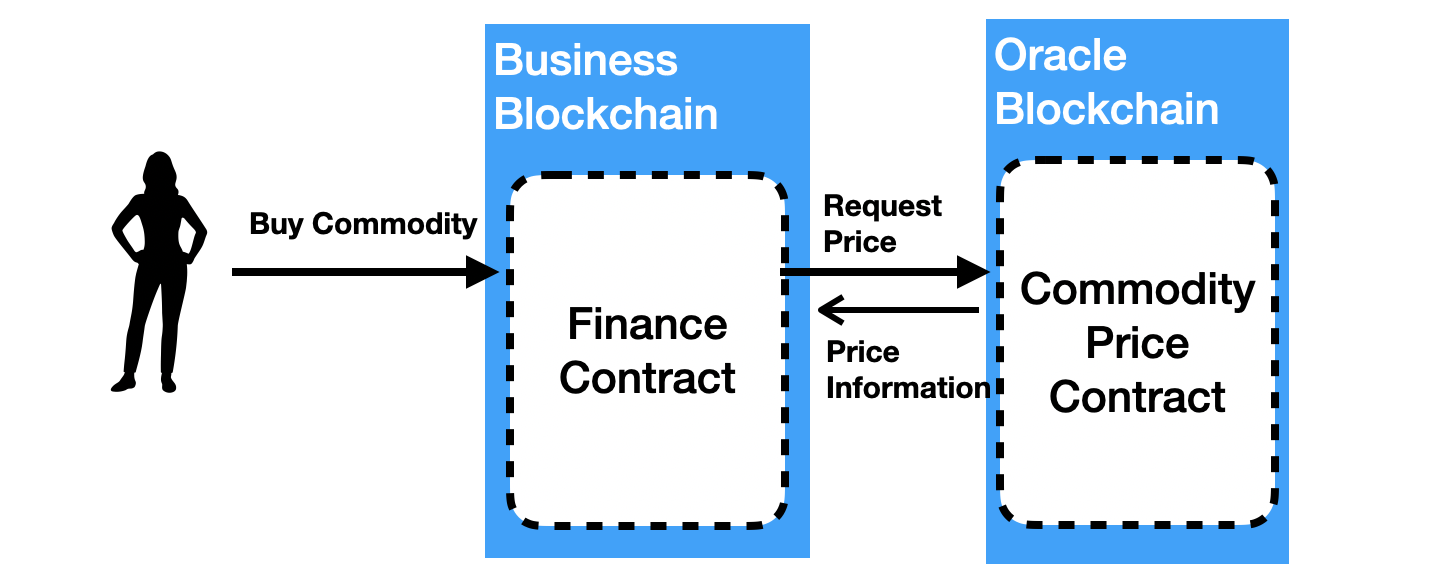}
\caption{Oracle Scenario}
\label{fig:oracle}
\end{figure}

An Oracle blockchain maintains a set of data that is valuable to other blockchain applications. For example, in Figure~\ref{fig:oracle} a company could publish commodity prices to their blockchain. They could charge other companies for the right to access their blockchain. The other companies could use cross-blockchain transactions to execute business logic on their own blockchain based on the information returned from the Oracle blockchain. 

\section{Experimental Setup}
\begin{figure} [h]     
\includegraphics[width=\columnwidth]{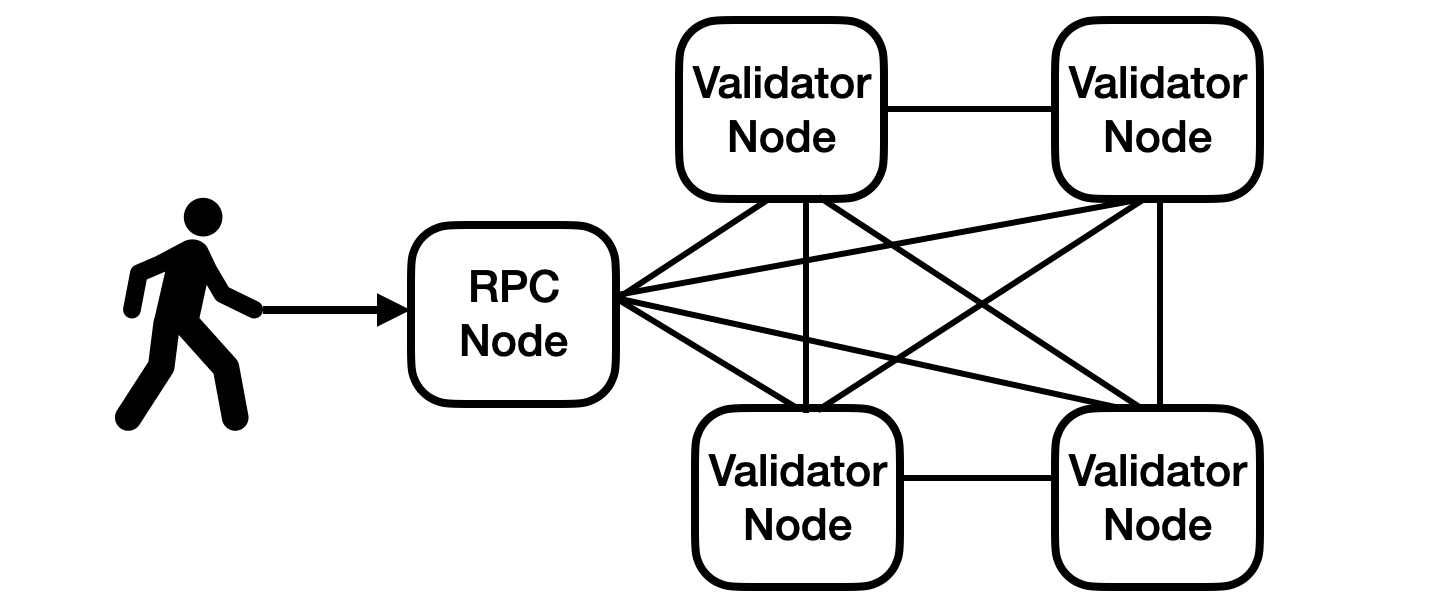}
\caption{Baseline Setup}
\label{fig:experiment}
\end{figure}

The performance of Hyperledger Besu version 1.4.4 with native crypto enabled, with an empty state trie, using IBFT2 consensus protocol, with no Byzantine nodes, when executing the \textit{open} function in the Hyperledger Caliper Benchmarks code Simple.sol \cite{caliper}, was measured using the configuration shown in Figure~\ref{fig:experiment}. All nodes were run on an Amazon Web Services c5d.4xlarge virtual machine (16 virtual CPUs, 32 GBytes of RAM, 450 GByte NVMe SSD, 2.25 Gbps elastic block storage bandwidth, 10 Gbps network connection). An application submitted transactions to a RPC Node that was connected with four validator nodes. All nodes were hosted in the same data centre.

In this configuration, the limiting factor for performance was the processing power of the CPU. For the purposes of defining base line performance, it will be assumed that code for each scenario executing in any function in a single blockchain transaction is able to execute at the measured rate of 375 transactions per second.

The LTACFC protocol requires multiple transactions to execute parts of the cross-blockchain function calls, as described in Section~\ref{sec:perf}.
To calculate the relative transaction rate, the  transaction rate of Hyperledger Besu will be scaled based on these numbers of transactions.

\section{Results}
This section analyses the differences in requirements for operating a node given the scenarios described in Section~\ref{sec:scenarios} for using standard and scalability variants of the LTACFC protocol, as compared to executing separate non-atomic single blockchain transactions. Table~\ref{table:scenarios} succinctly describes the scenarios in terms of cross-blockchain function call components. 

\begin{table}
  \centering
    \begin{tabular}{| l | l |}
    \hline
    Scenario & Description  \\
       \hline
       \hline
Hotel and Train &  Root with two Segments (updates) \\
       \hline
Supply Chain & Root with one Segment (update) \\
Provenance   &      \\
       \hline
Oracle           & Root with Segment (no updates) \\
       \hline
  \end{tabular}
  \caption{Analysis Scenarios}
  \label{table:scenarios}
\end{table}

Section ~\ref{sec:perf} explained that more transactions are needed on Root blockchains compared to Segment blockchains. This means that for the Supply Chain Provenance and Oracle scenarios, the Root blockchain is the blockchain that has the most overhead. The Root function call is on different blockchains for the Hotel and Train scenario, depending on which travel agency starts the booking. Table~\ref{table:processing1} shows the number of transactions needed for each scenario, with the Hotel and Train scenario split depending on whether a single travel agency makes all of the bookings and the Root blockchain is the bottle neck, or if a variety of travel agencies make bookings and the Segment blockchain is the bottle neck.

\begin{table}
  \centering
    \begin{tabular}{| l | l | l |}
    \hline
   \multirow{2}{*}{Scenario} &  \multicolumn{2}{|c|}{Number of Transactions}  \\
       \cline{2-3}
                         & LTACFC Protocol & LTACFC Protocol\\ 
                         &                             & Scalability \\ 
                         &               & Variant \\ 
       \hline
Hotel \& Train          & 1 (Start) & 1 (Start)  \\
(one Travel.        & +1 (Root) & +1 (Root)  \\
Agency)              & +1 (Clean) & +1 (Clean)  \\
                            & on Travel Agency & on Travel Agency \\
                            & blockchain           & blockchain \\
                            & +2 for block header & +1 for a root hash \\
                            & from Train blockchain & for each block \\
                            & +2 for block headers & \\
                            & from Hotel blockchain & \\
       \hline
Hotel \& Train               & 1 (Segment) & 1 (Segment)  \\
(many Travel          & +1 (Signal) & +1 (Signal)\\
 Agencies)             & on Hotel or Train  & on Hotel or Train \\
                               & blockchain            & blockchain \\
                                           & +2 for block headers & +1 for a root hash \\
                            & from Travel Agency       & for each block \\
                            & blockchains                   &     \\
       \hline
Supply Chain         & 1 (Start) & 1 (Start)  \\
Provenance          & +1 (Root) & +1 (Root)  \\
                            & +1 (Clean) & +1 (Clean)  \\
                            & on Supply Chain & on Supply Chain \\
                            & blockchain           & blockchain \\
                            & +2 for block headers & +1 for a root hash\\
                            & from Provenance  & for each block\\
                            & blockchain  & \\
       \hline
Oracle          & 1 (Start) & 1 (Start)  \\
                       & +1 (Root) & +1 (Root)  \\
                            & +1 (Clean) & +1 (Clean)  \\
                            & on Business & on Business \\
                            & blockchain           & blockchain \\
                            & +1 for a block header & +1 for a root hash \\
                            & from Oracle blockchain & for each block \\
       \hline
  \end{tabular}
  \caption{Number of Transactions on Blockchain that is Bottle Neck}
  \label{table:processing1}
\end{table}

\begin{table}
  \centering
    \begin{tabular}{| l | c | c |}
    \hline
   \multirow{2}{*}{Scenario} &  \multicolumn{2}{|c|}{Function Calls Per Second}  \\
       \cline{2-3}
                         & LTACFC Protocol & LTACFC Protocol\\ 
                         &                             & Scalability \\ 
                         &               & Variant \\ 
       \hline
Hotel and Train          & 53.5      & 124  \\
(one Travel.        & & \\
Agency)              &  &  \\
       \hline
Hotel and Train               & 93.7 & 186  \\
(many Travel          &   &  \\
 Agencies)             &    &   \\
       \hline
Supply Chain         & 75 & 124  \\
Provenance          &   &    \\
                            &   &    \\
       \hline
Oracle          & 93.75 & 124  \\
       \hline
  \end{tabular}
  \caption{Number of Function Calls Per Second based on Transaction Rate on Blockchain that is the Bottle Neck}
  \label{table:processing2}
\end{table}

Table~\ref{table:processing2} shows the expected cross-blockchain function call rate given the numbers of transactions that need to execute for various scenarios, given a base transaction rate of 375 tps, assuming a block period of one second, and assuming all transactions take the same amount of time to execute as the benchmark transactions. For example, for the Hotel and Train (many Travel Agencies) scenario using the scalability variant of the protocol, two transactions are needed, one for the Segment function call and one for the Signal function call. Additionally, one transaction is needed each second to transfer the root hash from the Block Header blockchain. Hence, the transaction rate is $375 \div 2 - 1$ = 186.5.

\section{Applicability to Ethereum 2 Cross-Shard Function Calls}
A simplified variant of the LTACFC protocol could be used for Ethereum 2 Cross-Shard function calls. Nodes on Ethereum 2 shards post root hashes called \textit{Crosslinks} to the Beacon Chain. The Crosslinks for all shards are available to all shards in the next block. This mechanism of cross-linking in combination with Ethereum 2's shared consensus mechanism replaces the cross-blockchain consensus mechanism described in Section~\ref{sec:consensus}. The rest of LTACFC protocol should be directly applicable as an Ethereum 2 Cross-Shard function call mechanism as described in this paper.

\section{Conclusion}
This paper introduces the LTACFC protocol, a protocol that provides atomic, synchronous, inter-contract function calls across blockchains, without the need to alter the blockchain platform software. No other technology provides the atomic capability and operates at Blockchain Layer 2. The protocol ensures updates across blockchains are either all committed or all ignored. A standard variant and a more scalable variant of the protocol are described. The standard variant results in a lower function call rate but lower latency, whereas the more scalable variant results in a far higher function call rate, but a higher latency. 

This complex functionality of the protocol comes with the cost of reduced blockchain system performance. For example, for the Hotel and Train scenario when there are many travel agencies and the standard variant of the protocol are used the function rate is 93.8 and 186 for the scalable protocol variant. This contrasts with 375 transactions per second for non-atomic single blockchain transactions. The LTACFC protocol numbers assume a block period of one second, and that all transactions take the same amount of time to execute as the benchmark transactions.

The performance of the LTACFC protocol depends on the size of the cross-blockchain call graph. Therefore architects utilising this technology need to understand their cross-blockchain functions call graph to optimise system performance. They need to determine important high value transactions that need to be cross-blockchain to minimise the impact on performance. 

This technique could be used for Ethereum 2 cross-shard function calls. In this shard-consensus scenario, the transactions required for cross-blockchain consensus would not be required. This means that this technique could be used with Ethereum 2 without the same degree of performance overhead as has been described in this paper.

\ifACM
\begin{acks}

\else

  \section*{Acknowledgment}

\fi

This research has been undertaken whilst we have been employed full-time at ConsenSys. Peter acknowledges the support of University of Queensland where he is completing his PhD, and in particular the support of his PhD supervisor Dr Marius Portmann. We acknowledge the researchers who have helped develop the Atomic Crosschain Transactions technology upon which this protocol is based: Dr Sandra Johnson, Dr David-Hyland Wood, Roberto Saltini, Horacio Mijail Anton Quiles, John Brainard, and Zhenyang Shi. We acknowledge Dr Catherine Jones, Dr Marius Portman, Dr David Hyland-Wood, Horacio Mijail Anton Quiles, and Dr Sandra Johnson for their thoughtful review comments and suggestions. 

\ifACM
\end{acks}

\bibliographystyle{ACM-Reference-Format}
\bibliography{ref}

\else

\bibliographystyle{IEEEtran}
\bibliography{IEEEabrv,ref}

\fi

\end{document}